\documentclass[twocolumn,prl,superscriptaddress,showpacs,preprintnumbers,amsmath,amssymb]{revtex4}

\usepackage{graphicx}
\usepackage{dcolumn}
\usepackage{bm}
\usepackage{color}

\begin{document}

\title{Dynamics of Nanometer-Scale Foil Targets Irradiated with Relativistically Intense Laser Pulses}

\author{R. H\"orlein}
    \email[Corresponding author: ]{rainer.hoerlein@mpq.mpg.de}
    \affiliation{Max-Planck-Institut f\"ur Quantenoptik, D-85748 Garching, Germany}
	\affiliation{Fakult\"at f\"ur Physik, Ludwig-Maximilians-Universit\"at M\"unchen, D-85748 Garching, Germany}

\author{S. Steinke}
	\affiliation{Max-Born-Institut, D-12489 Berlin, Germany}

\author{A. Henig}
\author{S. G. Rykovanov}
	\affiliation{Max-Planck-Institut f\"ur Quantenoptik, D-85748 Garching, Germany}
	\affiliation{Fakult\"at f\"ur Physik, Ludwig-Maximilians-Universit\"at M\"unchen, D-85748 Garching, Germany}

\author{M. Schn\"urer}
\author{T. Sokollik}
	\affiliation{Max-Born-Institut, D-12489 Berlin, Germany}

\author{D. Kiefer}
	\affiliation{Max-Planck-Institut f\"ur Quantenoptik, D-85748 Garching, Germany}
	\affiliation{Fakult\"at f\"ur Physik, Ludwig-Maximilians-Universit\"at M\"unchen, D-85748 Garching, Germany}

\author{D. Jung}
	\affiliation{Fakult\"at f\"ur Physik, Ludwig-Maximilians-Universit\"at M\"unchen, D-85748 Garching, Germany}
	\affiliation{Los Alamos National Laboratory, 87545 Los Alamos, NM, USA}

\author{X. Q. Yan}
	\affiliation{Max-Planck-Institut f\"ur Quantenoptik, D-85748 Garching, Germany}
	\affiliation{State Key Lab of Nuclear Physics and Technology, Peking University, 100871 Bejing, China}
	
%
\author{T. Tajima}
	\affiliation{Fakult\"at f\"ur Physik, Ludwig-Maximilians-Universit\"at M\"unchen, D-85748 Garching, Germany}
	\affiliation{Photomedical Research Center, JAEA, Kyoto, Japan}
	
\author{J. Schreiber}
	\affiliation{Max-Planck-Institut f\"ur Quantenoptik, D-85748 Garching, Germany}
	\affiliation{Fakult\"at f\"ur Physik, Ludwig-Maximilians-Universit\"at M\"unchen, D-85748 Garching, Germany}

\author{M. Hegelich}
	\affiliation{Fakult\"at f\"ur Physik, Ludwig-Maximilians-Universit\"at M\"unchen, D-85748 Garching, Germany}
	\affiliation{Los Alamos National Laboratory, 87545 Los Alamos, NM, USA}

\author{P. V. Nickles}
	\affiliation{Max-Born-Institut, D-12489 Berlin, Germany}
	\affiliation{Gwangju Institute of Science and Technology, GIST, Gwangju 500-712, Republic of Korea}

\author{M. Zepf}
	\affiliation{Max-Planck-Institut f\"ur Quantenoptik, D-85748 Garching, Germany}
	\affiliation{Department of Physics and Astronomy, Queens University Belfast, BT7 1NN Belfast, United Kingdom}

\author{G. D. Tsakiris}
	\affiliation{Max-Planck-Institut f\"ur Quantenoptik, D-85748 Garching, Germany}

\author{W. Sandner}
	\affiliation{Max-Born-Institut, D-12489 Berlin, Germany}

\author{D. Habs}
	\affiliation{Max-Planck-Institut f\"ur Quantenoptik, D-85748 Garching, Germany}
	\affiliation{Fakult\"at f\"ur Physik, Ludwig-Maximilians-Universit\"at M\"unchen, D-85748 Garching, Germany}

\date{\today}

\begin{abstract}
 	In this letter we report on an experimental study of high harmonic radiation generated in nanometer-scale foil targets irradiated under normal 
 	incidence. The experiments constitute the first unambiguous observation of odd-numbered relativistic harmonics generated by the 
 	$\vec{v}\times\vec{B}$ component of the Lorentz force verifying a long predicted property of solid target harmonics. Simultaneously the 
 	observed harmonic spectra allow in-situ extraction of the target density in an experimental scenario which is of utmost interest for applications 
 	such as ion acceleration by the radiation pressure of an ultraintense laser.   
\end{abstract}


\maketitle

		The interaction of ultra-intense laser pulses with nanometer-scale solid density foil targets has raised a lot of interest as it promises the 
		generation of monoenergetic ion and electron bunches via novel, more efficient acceleration mechanisms such as Radiation Pressure Acceleration 
		(RPA) \citep{Esi04Piston,Kli08RPA,Rob08RPA,Yan08PhaseStable,Hen09RPA}. The key to efficient RPA is to suppress heating of the target electrons 
		\citep{Kli08RPA,Rob08RPA} and careful control of the target surface areal density \citep{Yan08PhaseStable}. To obtain information on these 
		quantities, a diagnostic capable of probing the dynamics of the entire laser generated plasma during the acceleration process is necessary. We 
		show that Surface High Harmonic Generation (SHHG) in the transmission of nm-scale foil targets irradiated under normal incidence allow detailed 
		studies of crucial interaction parameters such as the target deformation and plasma density when the peak of the driving pulse interacts with the 
		target, i.e. while the particle acceleration takes place, without requiring an additional probe beam. Moreover our results advance the 
		understanding of the harmonic generation process itself as previous experiments have not addressed normal incidence interactions in detail. 
		
		The generation of high-harmonic radiation from solid density bulk 
		\citep{Bul94PhysPlasmas,Bae06RelSpikes,Tsa06NJP,Teu09RevSHHG,Nom08AC,Dro07KeV,Dro08Div,Que05CWE,Tar07ROM} as well as foil targets 
		\citep{Geo09Foils,Teu04RearsideHarm,Kru08Foils} has been studied extensively in recent years as it promises XUV and soft X-ray pulses of 
		attosecond duration with unprecedented intensities \citep{Tsa06NJP} and is a powerful probe of the ultrafast target dynamics. Recent experiments 
		have shown that the harmonics generated on solid targets are indeed phase-locked and emitted as a train of as-pulses \citep{Nom08AC} and have 
		demonstrated the efficient generation of harmonics up to the keV energy range \citep{Dro07KeV}. Simultaneously high harmonic generation gives 
		detailed insight into the laser plasma dynamics in high-intensity laser solid interactions allowing the probing of, for example, the plasma 
		density \citep{Que05CWE}, magnetic fields \citep{Tat02Mag}, surface dynamics \citep{Tar07ROM,Dro08Div} and electron heating \citep{Kru08Foils}. 
		
\begin{figure}[t]
	  	\includegraphics[width=0.44\textwidth]{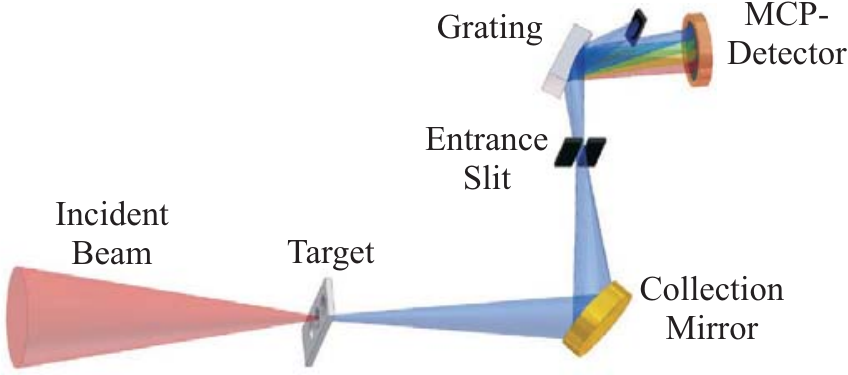} 		
			\caption{ Schematic drawing of the experimental setup.} 		
			\label{Setup}
		\end{figure}

		In this letter we present measurements of the harmonic emission in transmission of nm-scale (sub-20~nm) high-density ($\approx$ 2.7~g\,cm$^{-3}$) 
		diamond-like carbon (DLC) foils irradiated with ultra-high contrast linearly polarized (LP) laser pulses at normal incidence [Fig.~\ref{Setup}]. 
		We show that the harmonic emission, apart from being an interesting radiation source, is a versatile probe of the ultrafast target dynamics. The 
		experiments give insight into the three dimensional nature of the laser-foil interaction dynamics and allow the determination of the target 
		density at the instant when the peak of the driving pulse interacts with it. This constitutes the first measurement of critical plasma 
		parameters in normal incidence interactions of relativistic laser pulses with nm-scale foil targets relevant, for example, for ion acceleration 
		e.g. \citep{Ste10Ions}. Moreover the experiments confirm a central theoretical prediction in SHHG at normal incidence 
		\citep{Bul94PhysPlasmas,Bae06RelSpikes,Tsa06NJP} that has so far lacked conclusive experimental proof. In this case SHHG should 
		be dominated by the $\vec{v}\times\vec{B}$ component of the Lorentz force oscillating at twice the laser frequency. Unlike in the oblique 
		incidence case this should result in the generation of only odd-numbered harmonics.

		Like in the interaction with bulk targets two distinct mechanisms capable of generating high-order harmonics in the transmission of nm-scale foils 
		have been identified \citep{Geo09Foils}. For sub- and weakly relativistic intensities corresponding to a normalized vector potential 
		$a_0=\sqrt{I_\mathrm{L}\lambda_\mathrm{L}^2/(1.37\times10^{18}Wcm^{-2}\mu m^{2})}$ below or on the order of unity, where $I_\mathrm{L}$ and 
		$\lambda_\mathrm{L}$ stand for the cycle-averaged intensity and the wavelength of the incident laser light, harmonics are generated via linear 
		mode-conversion of plasma waves \citep{She05LMC} in the density gradient on the rear side of the foil. These plasma waves, unlike on the front 
		side of the target, are excited indirectly by as-electron bursts propagating up the rear side density gradient. These bunches are accelerated in 
		the electric field set up by other electron bunches generated on the front side that penetrate through the foil 
		\citep{Geo09Foils,Teu04RearsideHarm}. This mechanism is conventionally called Coherent Wake Emission (CWE) \citep{Que05CWE}. For larger 
		intensities and especially in the relativistic limit ($a_0\gg~1$) harmonic radiation can also be emitted by plasma electrons close to the critical 
		density layer coherently oscillating with velocities close to the speed of light in the driving laser field 
		\citep{Bul94PhysPlasmas,Bae06RelSpikes,Tsa06NJP}. In the case of very thin foil targets this mechanism can also generate harmonics emitted in the 
		forward direction that propagate through the target and can be observed at the rear side \citep{Geo09Foils,Kru08Foils}. The predominant mechanism 
		in a specific experiment can be determined by the characteristics of the detected harmonic spectrum. While relativistic harmonics exhibit an 
		intensity dependent spectral cutoff, CWE harmonics do not. Instead, they have a distinct high energy cutoff determined by the maximum plasma 
		frequency and thus peak density in the target. In this case the highest generated harmonic order is 
		$q_\mathrm{co}=\omega_\mathrm{p,max}/\omega_\mathrm{L}$ where $\omega_\mathrm{L}$ and $\omega_\mathrm{p,max}=\sqrt{n_\mathrm{e,max}e^2/\epsilon_0 
		m_e}$ with $n_\mathrm{e,max}$ the peak electron density are the laser and peak plasma frequency respectively. In addition, the generation of CWE 
		harmonics requires oblique incidence as the conditions for mode conversion cannot be fulfilled otherwise \citep{She05LMC}. In contrast 
		relativistic harmonics can also be generated under normal incidence in which case the electron oscillations are driven by the 
		$\vec{v}\times\vec{B}$ component of the driving force oscillating at twice the laser frequency resulting in the generation of only odd harmonics 
		\citep{Tsa06NJP}.
			
		In the case of a mildly relativistic laser pulse incident normally onto a thin foil the harmonic spectrum may show signatures of both harmonic 
		generation mechanisms, which one dominates will depend on the detailed dynamics of the interaction. Especially denting of the target in the focus 
		will alter the interaction significantly as this results in effectively oblique incidence on the sides of the focal spot. Due to the lower 
		intensity level in these regions of the focus the harmonic generation will likely be dominated by CWE in such regions resulting in a spectrum 
		containing all harmonics with a density dependent cutoff. If relativistic harmonics are also generated they will originate predominantly from the 
		center of the focus where the intensity is highest and likely display only odd orders owing to the true normal incidence in this region. The 
		relative efficiency of the two mechanisms will vary depending on how planar the laser target interaction is, resulting in different intensity 
		ratios between odd and even harmonics. In addition target non-uniformities across the focal region can also influence the effective angle of 
		incidence of the driving electric field. 
	
\begin{figure}[t]
		  	\includegraphics[width=0.44\textwidth]{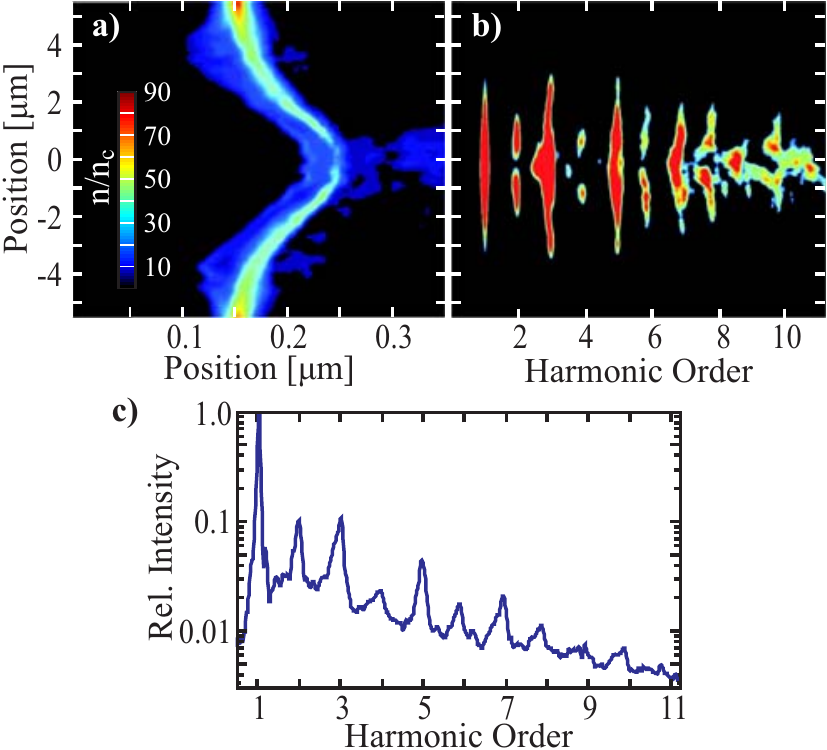} 		
				\caption{	Results of 2D PIC-simulations of the interaction of a relativistic laser pulse with a triangular shaped target with 25~nm gradients 
									on each side and a peak density of 100n$_c$. Plots (a) show the electron density distribution at the instance when the peak of the 
									driving pulse arrives at the initial target position (note the different length scales used to emphasize the denting of the target) 
									and (b) the time integrated harmonic spectra emitted in the forward direction as a function of radial position recorded at a 
									position just behind the target surface. A clear difference in the emission characteristics on and off-axis is visible. Radially 
									integrating the individual spectra in (b) to account for the collection of the signal with a mirror yields the spectrum shown in (c) 
									exhibiting all harmonics with an enhancement of the odd orders.} 	
				\label{Simulations}
			\end{figure}

		To verify this argumentation we have conducted 2D Particle-In-Cell (PIC) simulations of the laser foil interaction at normal incidence using the 
		code PICWIG \citep{Ryk08PolGat}. As will be discussed later even for the ultra-clean laser pulses employed in this experiment the target will have 
		expanded in the rising edge of the 45~fs pulse with $a_0=3.6$. Taking this into account the simulations were initialized using a triangular 
		density profile with a peak density of $n_0=100n_c$ and a linear ramp of length $25~nm$ on both the front and the rear side. Considering that the 
		original target density in the experiment is approximately 480n$_c$ this corresponds to a solid density foil of approximately 5~nm original 
		thickness. The laser pulse was Gaussian both in space and time (spot size $3~\mu m$ FWHM, duration duration 15 cycles FWHM in the field) and 
		incident normally onto the target. The size of the simulation box is $6~\lambda_{\mathrm{L}}$ in laser propagation direction and 
		$15~\lambda_{\mathrm{L}}$ in polarization direction. The time step is $\tau_{\mathrm{L}}/400$ and the laser propagation direction spatial step 
		correspondingly is $\lambda_{\mathrm{L}}/400$ where $\tau_{\mathrm{L}}$ is the period of the driving laser. The results of the simulation are 
		depicted in Fig.~\ref{Simulations}. The denting of the foil during the interaction is clearly visible in Fig.~\ref{Simulations}(a) where the 
		electron density near the instance when the peak of the pulse interacts with the target is shown. Fig.~\ref{Simulations}(b) shows the harmonic 
		spectrum emitted in the forward direction as a function of position across the target recorded at a position just behind the target surface. 
		On-axis only odd harmonics are generated which can only originate from relativistic electron oscillations as the condition for mode conversion 
		necessary for CWE \citep{She05LMC} cannot be fulfilled in this geometry while off-axis all harmonics are generated owing to the effectively 
		oblique incidence of the driving pulse. To derive the total harmonic spectrum collected in the experiment after propagation to the detector we sum 
		over the spectra emitted at the different positions in the focus. The resulting power spectrum is shown in Fig.~\ref{Simulations}(c). The 
		integrated spectrum contains all harmonic orders but an enhancement of odd harmonics originating from the center of the generation region is 
		visible constituting a clear signature of relativistic harmonic generation on axis.  
	
		The experiments presented in this letter were conducted using the 30~TW laser facility at the Max-Born-Institute in Berlin. The Ti:sapphire laser 
		system delivered pulses with an energy of 0.7~J and a pulse duration of 45~fs at a central wavelength of 810~nm to the target. To enhance the 
		temporal contrast of the laser a recollimating double plasma mirror (DPM) \citep{And09PM} was introduced into the system enhancing the temporal 
		contrast to better than 1:10$^{10}$ on the few picosecond scale. The beam was focused to a near diffraction-limited focal spot of 3.6~$\mu$m FWHM 
		diameter using a dielectrically coated f/2.5 off-axis parabola resulting in a peak focused intensity of 	
		$I_{0\mathrm{peak}}=5\times10^{19}~W/cm^{2}$ corresponding to a normalized vector potential of $a_{0\mathrm{peak}}\approx 5$.
	
		Free-standing DLC foils like the ones used in \citep{Hen09RPA,Ste10Ions} ranging from approximately 5 to 17~nm in thickness were irradiated under 
		normal incidence in the focus of the driving laser beam. The radiation emitted in the laser propagation direction was collected using a spherical 
		mirror with an unprotected gold coating positioned under an angle of 45 degrees [see Fig.~\ref{Setup}]. The resulting line focus was 
		placed on the entrance slit of a normal incidence ACTON VM-502 VUV-spectrometer equipped with a Micro-Channel-Plate (MCP) detector and a 
		fiber-coupled CCD-camera. The spectrometer allowed the detection of radiation from 140 down to 50~nm corresponding to harmonics 7 to 16 of the 
		fundamental laser wavelength.
	
			\begin{figure}[t]
			  	\includegraphics[width=0.44\textwidth]{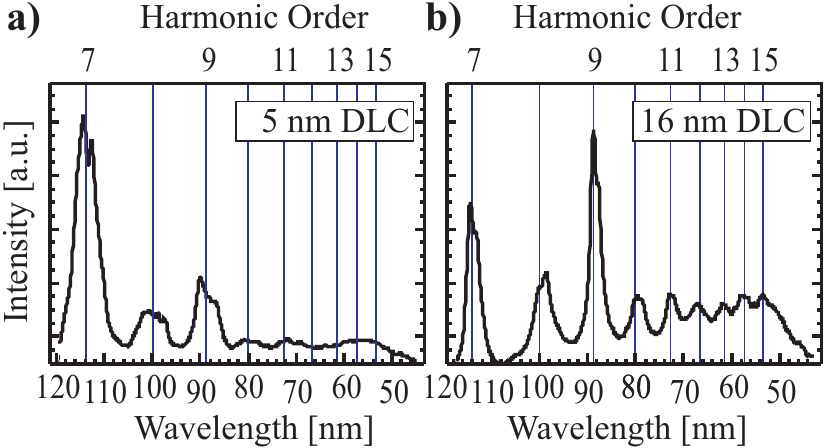} 		
					\caption{	Normalized high harmonic spectra obtained from targets with two different thicknesses.} 		
					\label{Spectra}
			\end{figure}
	
		Two typical normalized harmonic spectra obtained from targets of different thicknesses are depicted in Fig.~\ref{Spectra}. The absolute peak 
		intensity in (a) is approximately twice as big as in (b). Both spectra show odd and even harmonics with a pronounced enhancement of the odd 
		harmonics of orders 7 (H7) and 9 (H9). The spectra do however differ significantly in the highest harmonic visible. The spectrum from the thinner 
		target [Fig.~\ref{Spectra}(a)] shows harmonics up to H9 whereas radiation with significantly shorter wavelength up to H15 is generated from the 
		thicker target [Fig.~\ref{Spectra}(b)].
	
\begin{figure}[t]
			 	\includegraphics[width=0.44\textwidth]{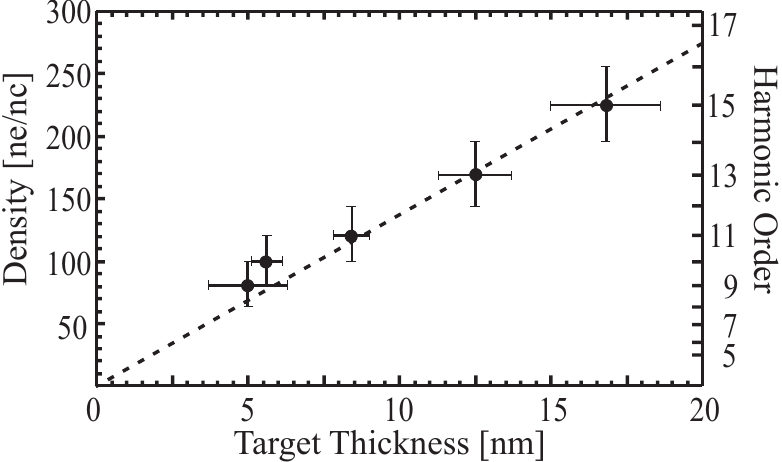} 		
				\caption{	Peak target density and cutoff harmonic plotted as a function of original target thickness. The dashed line corresponds to the peak 
									density expected for a one dimensional foil expansion with a scale length of 18~nm on each side. The spectra shown in Fig. 
									\ref{Spectra}(a) and (b) correspond (from left to right) to the second and the fifth data point respectively.}		
				\label{Density}
			\end{figure}

		These spectral properties suggest that the harmonics in our experiment are indeed generated by two different mechanisms as predicted by 
		simulations, one generating predominantly odd and one producing all harmonics. The thickness dependent cutoff with higher harmonics from thicker 
		foils in combination with the moderate intensities on the sides of the focus suggest that this part of the spectrum is generated by CWE 
		\citep{Geo09Foils}. This means that the target has to be dented significantly to facilitate motion of the plasma electrons in and out of the 
		surface in the electric field as predicted by the simulation. The trend of higher harmonics from thicker targets is visible for all measured 
		targets and is displayed in Fig.~\ref{Density}, where the cutoff harmonic order and corresponding peak target density is plotted versus the 
		original foil thickness. The peak density in the interaction region as inferred from the harmonic generation emitted during the most intense phase 
		of the laser foil interaction, is found to increase linearly with initial target thickness. This suggests that the target plasma expands with 
		similar velocities in all cases resulting in densities during the interaction that are significantly lower than those of the original foil. For 
		all studied targets the decrease in density is consistent with an exponential density ramp with a scale length of approximately 18~nm on each side 
		of the foil at the onset of the relativistic interaction. The dashed line in Fig.~\ref{Density} shows the peak target densities expected in such a 
		scenario and is in very good agreement with the measured data. To check these values for consistency we estimate the expansion of the foil after 
		ionization and prior to the relativistic interaction considering a gaussian pulse shape on the rising edge of the 45~fs pulse. Assuming uniform 
		energy deposition in the foil \citep{Pri95Absorption} the ion sound velocity rapidly increases over a time window of 70~fs to approximately 
		$c_{\mathrm{s,peak}}=5\times 10^7~cm/s$ corresponding to a hot electron temperature $T_e\approx5~keV$ at which the plasma becomes non-collisional 
		under our experimental conditions \citep{Kruer,Gibbon,Bla96Expansion}. During this time the average sound velocity is found to be 
		$c_{\mathrm{s,mean}}\approx2.2\times 10^7~cm/s$ which results in an expansion by 15.4~nm in good agreement with the measured data. Thus, in 
		addition to the determination of the target density, the experiment gives information about the target heating prior to the relativistic 
		interaction and demonstrates a fundamental limitation of the nm-foil density at the instance of the interaction with the peak of the laser pulse 
		which can be crucial when even thinner targets are employed.
	
		In conclusion we have presented the first experiment demonstrating the generation of high order harmonics in the transmission of few-nm-scale foil 
		targets irradiated at normal incidence. The measurements in combination with two dimensional PIC-simulations demonstrate that the harmonics are 
		generated via two different mechanisms in different regions of the laser focus giving detailed insight into the dynamics of the foil target under 
		experimental conditions also of interest for example for particle acceleration experiments. While non-uniformities such as denting of the target 
		leads to effectively oblique incidence of the driving field on the sides of the focus and the generation of all harmonics in this region, only odd 
		harmonics are generated exactly on the laser axis. This constitutes the first unambiguous demonstration of relativistic harmonic generation at 		
		normal incidence as predicted in many theoretical publications. Harmonics are mainly generated via CWE in regions of oblique incidence which 
		allows the determination of the instantaneous target density of the foil in the relativistic interaction. The densities inferred from the observed 
		harmonic spectra are in good agreement with the one dimensional expansion of the foil targets. This expansion occurs even for perfect gaussian 
		laser pulses and thus imposes a fundamental limit on the peak density of few-nm-scale foil targets during the relativistic part of the laser solid 
		interaction. The experiments demonstrate that, beyond the fundamental study of the harmonic generation process itself, this method is a powerful 
		diagnostic of the laser plasma interaction that can be employed in many experimental scenarios pertaining to efficient laser particle 
		acceleration including RPA with circular polarization. In fact, the method does not even require any modifications to existing particle 
		acceleration experiments except for the implementation of a pickoff optic as all the information is generated by the driving laser beam itself.
		
		We would like to thank the Berlin laser staff for their support. This work was funded in part by the DFG through SFB Transregio18 and the Cluster 
		of Excellence Munich Center for Advanced Photonics (MAP) and by the Association EURATOM--Max-Planck-Institut f\"ur Plasmaphysik. A. H., S. G. R., 
		D. K. and D. J. acknowledge financial support from IMPRS-APS. X. Q. Y. acknowledges financial support from the Humboldt Foundation and NSFC 
		(10935002).


\begin{thebibliography}{0}
\expandafter\ifx\csname natexlab\endcsname\relax\def\natexlab#1{#1}\fi
\expandafter\ifx\csname bibnamefont\endcsname\relax
  \def\bibnamefont#1{#1}\fi
\expandafter\ifx\csname bibfnamefont\endcsname\relax
  \def\bibfnamefont#1{#1}\fi
\expandafter\ifx\csname citenamefont\endcsname\relax
  \def\citenamefont#1{#1}\fi
\expandafter\ifx\csname url\endcsname\relax
  \def\url#1{\texttt{#1}}\fi
\expandafter\ifx\csname urlprefix\endcsname\relax\def\urlprefix{URL }\fi
\providecommand{\bibinfo}[2]{#2}
\providecommand{\eprint}[2][]{\url{#2}}

\end{thebibliography}


\begin{thebibliography}{26}
	
		\expandafter\ifx\csname natexlab\endcsname\relax\def\natexlab#1{#1}\fi
		\expandafter\ifx\csname bibnamefont\endcsname\relax
		  \def\bibnamefont#1{#1}\fi
		\expandafter\ifx\csname bibfnamefont\endcsname\relax
		  \def\bibfnamefont#1{#1}\fi
		\expandafter\ifx\csname citenamefont\endcsname\relax
		  \def\citenamefont#1{#1}\fi
		\expandafter\ifx\csname url\endcsname\relax
		  \def\url#1{\texttt{#1}}\fi
		\expandafter\ifx\csname urlprefix\endcsname\relax\def\urlprefix{URL }\fi
		\providecommand{\bibinfo}[2]{#2}
		\providecommand{\eprint}[2][]{\url{#2}}

		\bibitem[{\citenamefont{Esirkepov et~al.}(2004)\citenamefont{Esirkepov,
		  Borghesi, Bulanov, Mourou, and Tajima}}]{Esi04Piston}
		\bibinfo{author}{\bibfnamefont{T.}~\bibnamefont{Esirkepov}} \textit{et al.},
		  \bibinfo{journal}{Phys. Rev. Lett.} \textbf{\bibinfo{volume}{92}},
		  \bibinfo{pages}{175003} (\bibinfo{year}{2004}).
		
		\bibitem[{\citenamefont{Klimo et~al.}(2008)\citenamefont{Klimo, Psikal,
		  Limpouch, and Tikhonchuk}}]{Kli08RPA}
		\bibinfo{author}{\bibfnamefont{O.}~\bibnamefont{Klimo}},
		  \bibinfo{author}{\bibfnamefont{J.}~\bibnamefont{Psikal}},
		  \bibinfo{author}{\bibfnamefont{J.}~\bibnamefont{Limpouch}}, \bibnamefont{and}
		  \bibinfo{author}{\bibfnamefont{V.~T.} \bibnamefont{Tikhonchuk}},
		  \bibinfo{journal}{Phys. Rev. ST Accel. Beams} \textbf{\bibinfo{volume}{11}},
		  \bibinfo{pages}{031301} (\bibinfo{year}{2008}).
		
		\bibitem[{\citenamefont{Robinson et~al.}(2008)\citenamefont{Robinson, Zepf,
		  Kar, Evans, and Bellei}}]{Rob08RPA}
		\bibinfo{author}{\bibfnamefont{A.~P.~L.} \bibnamefont{Robinson}} \textit{et al.},
		  \bibinfo{journal}{New J. Phys.} \textbf{\bibinfo{volume}{10}},
		  \bibinfo{pages}{013021} (\bibinfo{year}{2008}).
		
		\bibitem[{\citenamefont{Yan et~al.}(2008)\citenamefont{Yan, Lin, Sheng, Guo,
		  Liu, Lu, Fang, and Chen}}]{Yan08PhaseStable}
		\bibinfo{author}{\bibfnamefont{X.~Q.} \bibnamefont{Yan}} \textit{et al.},%
		  \bibinfo{journal}{Phys. Rev. Lett.} \textbf{\bibinfo{volume}{100}},
		  \bibinfo{pages}{135003} (\bibinfo{year}{2008}).
		
		\bibitem[{\citenamefont{Henig et~al.}(2009)\citenamefont{Henig, Steinke,
		  Schn\"urer, Sokollik, H\"orlein, Kiefer, Jung, Schreiber, Hegelich, Yan
		  et~al.}}]{Hen09RPA}
		\bibinfo{author}{\bibfnamefont{A.}~\bibnamefont{Henig}} \textit{et al.},
		  \bibinfo{journal}{Phys. Rev. Lett.}
		  \textbf{\bibinfo{volume}{103}}, \bibinfo{pages}{245003}
		  (\bibinfo{year}{2009}).
		
			\bibinfo{journal}{Eur. Phys. J. D}
		  \textbf{\bibinfo{volume}{55}}, \bibinfo{pages}{427} (\bibinfo{year}{2009}).
		
		\bibitem[{\citenamefont{Bulanov et~al.}(1994)\citenamefont{Bulanov, Naumova,
		  and Pegoraro}}]{Bul94PhysPlasmas}
		\bibinfo{author}{\bibfnamefont{S.~V.} \bibnamefont{Bulanov}},
		  \bibinfo{author}{\bibfnamefont{N.~M.} \bibnamefont{Naumova}},
		  \bibnamefont{and} \bibinfo{author}{\bibfnamefont{F.}~\bibnamefont{Pegoraro}},
		  \bibinfo{journal}{Phys. Plasmas} \textbf{\bibinfo{volume}{1}},
		  \bibinfo{pages}{745} (\bibinfo{year}{1994}).
		
		\bibitem[{\citenamefont{Baeva et~al.}(2006)\citenamefont{Baeva, Gordienko, and
		  Pukhov}}]{Bae06RelSpikes}
		\bibinfo{author}{\bibfnamefont{T.}~\bibnamefont{Baeva}},
		  \bibinfo{author}{\bibfnamefont{S.}~\bibnamefont{Gordienko}},
		  \bibnamefont{and} \bibinfo{author}{\bibfnamefont{A.}~\bibnamefont{Pukhov}},
		  \bibinfo{journal}{Phys. Rev. E} \textbf{\bibinfo{volume}{74}},
		  \bibinfo{pages}{046404} (\bibinfo{year}{2006}).
		
		\bibitem[{\citenamefont{Tsakiris et~al.}(2006)\citenamefont{Tsakiris, Eidmann,
		  Meyer-ter Vehn, and Krausz}}]{Tsa06NJP}
		\bibinfo{author}{\bibfnamefont{G.~D.} \bibnamefont{Tsakiris}},
		  \bibinfo{author}{\bibfnamefont{K.}~\bibnamefont{Eidmann}},
		  \bibinfo{author}{\bibfnamefont{J.}~\bibnamefont{Meyer-ter Vehn}},
		  \bibnamefont{and} \bibinfo{author}{\bibfnamefont{F.}~\bibnamefont{Krausz}},
		  \bibinfo{journal}{New J. Phys.} \textbf{\bibinfo{volume}{8}},
		  \bibinfo{pages}{19} (\bibinfo{year}{2006}).
		
		\bibitem[{\citenamefont{Teubner and Gibbon}(2009)}]{Teu09RevSHHG}
		\bibinfo{author}{\bibfnamefont{U.}~\bibnamefont{Teubner}} \bibnamefont{and}
		  \bibinfo{author}{\bibfnamefont{P.}~\bibnamefont{Gibbon}},
		  \bibinfo{journal}{Rev. Mod. Phys.} \textbf{\bibinfo{volume}{81}},
		  \bibinfo{pages}{445} (\bibinfo{year}{2009}).
		
		\bibitem[{\citenamefont{Nomura et~al.}(2009)\citenamefont{Nomura, H\"orlein,
		  Tzallas, Dromey, Rykovanov, Major, Osterhoff, Karsch, Veisz, Zepf
		  et~al.}}]{Nom08AC}
		\bibinfo{author}{\bibfnamefont{Y.}~\bibnamefont{Nomura}} \textit{et al.},
		  \bibinfo{journal}{Nat. Phys.} \textbf{\bibinfo{volume}{5}},
		  \bibinfo{pages}{124} (\bibinfo{year}{2009}).
		
		\bibitem[{\citenamefont{Dromey et~al.}(2007)\citenamefont{Dromey, Kar, Bellei,
		  Carroll, Clarke, Green, Kneip, Markey, Nagel, Simpson et~al.}}]{Dro07KeV}
		\bibinfo{author}{\bibfnamefont{B.}~\bibnamefont{Dromey}} \textit{et al.},
			\bibinfo{journal}{Phys. Rev. Lett.}
		  \textbf{\bibinfo{volume}{99}}, \bibinfo{pages}{085001}
		  (\bibinfo{year}{2007}).
		
		\bibitem[{\citenamefont{Dromey et~al.}(2009)\citenamefont{Dromey, Adams,
		  H\"orlein, Nomura, Rykovanov, Caroll, Foster, Kar, Markey, McKenna
		  et~al.}}]{Dro08Div}
		\bibinfo{author}{\bibfnamefont{B.}~\bibnamefont{Dromey}} \textit{et al.},
		  \bibinfo{journal}{Nat. Phys.}
		  \textbf{\bibinfo{volume}{5}}, \bibinfo{pages}{146} (\bibinfo{year}{2009}).
		
		\bibitem[{\citenamefont{Qu\'er\'e et~al.}(2006)\citenamefont{Qu\'er\'e, Thaury,
		  Monot, Dobosz, Martin, Geindre, and Audebert}}]{Que05CWE}
		\bibinfo{author}{\bibfnamefont{F.}~\bibnamefont{Qu\'er\'e}} \textit{et al.},
		  \bibinfo{journal}{Phys. Rev. Lett.} \textbf{\bibinfo{volume}{96}},
		  \bibinfo{pages}{125004} (\bibinfo{year}{2006}).
	%
		
		\bibitem[{\citenamefont{Tarasevitch et~al.}(2007)\citenamefont{Tarasevitch,
		  Lobov, Wuensche, and von~der Linde}}]{Tar07ROM}
		\bibinfo{author}{\bibfnamefont{A.}~\bibnamefont{Tarasevitch}},
		  \bibinfo{author}{\bibfnamefont{K.}~\bibnamefont{Lobov}},
		  \bibinfo{author}{\bibfnamefont{C.}~\bibnamefont{Wuensche}}, \bibnamefont{and}
		  \bibinfo{author}{\bibfnamefont{D.}~\bibnamefont{von~der Linde}},
		  \bibinfo{journal}{Phys. Rev. Lett.} \textbf{\bibinfo{volume}{98}},
		  \bibinfo{pages}{103902} (\bibinfo{year}{2007}).
		
		\bibitem[{\citenamefont{George et~al.}(2009)\citenamefont{George, Qu\'er\'e,
		  Thaury, Bonnaud, and Martin}}]{Geo09Foils}
		\bibinfo{author}{\bibfnamefont{H.}~\bibnamefont{George}} \textit{et al.},
		  \bibinfo{journal}{New J. Phys.} \textbf{\bibinfo{volume}{11}},
		  \bibinfo{pages}{113028} (\bibinfo{year}{2009}).
		
		\bibitem[{\citenamefont{Teubner et~al.}(2004)\citenamefont{Teubner, Eidmann,
		  Wagner, Andiel, Pisani, Tsakiris, Witte, Meyer-ter Vehn, Schlegel, and
		  F\"orster}}]{Teu04RearsideHarm}
		\bibinfo{author}{\bibfnamefont{U.}~\bibnamefont{Teubner}} \textit{et al.},
		  \bibinfo{journal}{Phys. Rev. Lett.} \textbf{\bibinfo{volume}{92}},
		  \bibinfo{pages}{185001} (\bibinfo{year}{2004}).
		
		\bibitem[{\citenamefont{Krushelnick et~al.}(2008)\citenamefont{Krushelnick,
		  Rozmus, Wagner, Beg, Bochkarev, Clark, Dangor, Evans, Gopal, Habara
		  et~al.}}]{Kru08Foils}
		\bibinfo{author}{\bibfnamefont{K.}~\bibnamefont{Krushelnick}} \textit{et al.},
		  \bibinfo{journal}{Phys. Rev. Lett.}
		  \textbf{\bibinfo{volume}{100}}, \bibinfo{pages}{125005}
		  (\bibinfo{year}{2008}).
		  
		\bibitem[{\citenamefont{Tatarakis et~al.}(2002)\citenamefont{Tatarakis, Watts,
		  Beg, Clark, Dangor, Gopal, Haines, Norreys, Wagner, Wei et~al.}}]{Tat02Mag}
		\bibinfo{author}{\bibfnamefont{M.}~\bibnamefont{Tatarakis}} \textit{et al.},
		  \bibinfo{journal}{Nature}
		  \textbf{\bibinfo{volume}{415}}, \bibinfo{pages}{280} (\bibinfo{year}{2002}).

	\bibitem{Ste10Ions}
			\bibinfo{author}{\bibfnamefont{S.}~\bibnamefont{Steinke}} \textit{et al.},
		  \bibinfo{journal}{Laser Part. Beams} \textbf{\bibinfo{volume}{28}},
		  \bibinfo{pages}{215} (\bibinfo{year}{2010}).
		
		\bibitem[{\citenamefont{Sheng et~al.}(2005)\citenamefont{Sheng, Mima, Zhang,
		  and Sanuki}}]{She05LMC}
		\bibinfo{author}{\bibfnamefont{Z.-M.} \bibnamefont{Sheng}},
		  \bibinfo{author}{\bibfnamefont{K.}~\bibnamefont{Mima}},
		  \bibinfo{author}{\bibfnamefont{J.}~\bibnamefont{Zhang}}, \bibnamefont{and}
		  \bibinfo{author}{\bibfnamefont{H.}~\bibnamefont{Sanuki}},
		  \bibinfo{journal}{Phys. Rev. Lett.} \textbf{\bibinfo{volume}{94}},
		  \bibinfo{pages}{095003} (\bibinfo{year}{2005}).
		
		\bibitem[{\citenamefont{Rykovanov et~al.}(2008)\citenamefont{Rykovanov,
		  Geissler, Meyer-ter Vehn, and Tsakiris}}]{Ryk08PolGat}
		\bibinfo{author}{\bibfnamefont{S.}~\bibnamefont{Rykovanov}},
		  \bibinfo{author}{\bibfnamefont{M.}~\bibnamefont{Geissler}},
		  \bibinfo{author}{\bibfnamefont{J.}~\bibnamefont{Meyer-ter Vehn}},
		  \bibnamefont{and} \bibinfo{author}{\bibfnamefont{G.~D.}
		  \bibnamefont{Tsakiris}}, \bibinfo{journal}{New J. Phys.}
		  \textbf{\bibinfo{volume}{10}}, \bibinfo{pages}{025025}
		  (\bibinfo{year}{2008}).
		
		\bibitem{And09PM}
		\bibinfo{author}{\bibfnamefont{A. A.}~\bibnamefont{Andreev}} \textit{et al.},
		  \bibinfo{journal}{Phys. Plasmas} \textbf{\bibinfo{volume}{16}},
		  \bibinfo{pages}{013103} (\bibinfo{year}{2009}).
		
			\bibitem{Pri95Absorption}
			\bibinfo{author}{\bibfnamefont{D. F.}~\bibnamefont{Price}} \textit{et al.},
		  \bibinfo{journal}{Phys. Rev. Lett.} \textbf{\bibinfo{volume}{75}},
		  \bibinfo{pages}{252} (\bibinfo{year}{1995}).
		
		\bibitem[{\citenamefont{Kruer}(1988)}]{Kruer}
		\bibinfo{author}{\bibfnamefont{W.~L.} \bibnamefont{Kruer}},
		  \emph{\bibinfo{title}{The Physics of Laser Plasma Interactions}}
		  (\bibinfo{publisher}{Addison-Wesley}, \bibinfo{address}{Redwood City, CA},
		  \bibinfo{year}{1988}).
		
		\bibitem[{\citenamefont{Gibbon}(2005)}]{Gibbon}
		\bibinfo{author}{\bibfnamefont{P.}~\bibnamefont{Gibbon}},
		  \emph{\bibinfo{title}{Short Pulse Laser Interaction with Matter}}
		  (\bibinfo{publisher}{Imperial College Press}, \bibinfo{address}{London},
		  \bibinfo{year}{2005}).
		
		\bibitem[{\citenamefont{Blanc et~al.}(1996)\citenamefont{Blanc, Audebert,
		  Falli\`{e}s, Geindre, Gauthier, Santos, Mysyrowicz, and
		  Antonetti}}]{Bla96Expansion}
		\bibinfo{author}{\bibfnamefont{P.}~\bibnamefont{Blanc}} \textit{et al.},
		  \bibinfo{journal}{J. Opt. Soc. Am. B} \textbf{\bibinfo{volume}{13}},
		  \bibinfo{pages}{118} (\bibinfo{year}{1996}).
		  

	\end{thebibliography}
\end{document}